\documentclass{appbav}  
\usepackage{epsfig}
\usepackage{cite}

\def\titleheading{\small\rm DESY 05-\desyNo, \small\rm DCPT/05/\dcptNo, 
 IPPP/05/\ipppNo \hfil hep-ph/\hepNo}

\def\desyNo{226}
\def\dcptNo{140}
\def\ipppNo{70}
\def\hepNo{0511113}

\newcommand{\hspn}{{\hspace{-5mm}}}

\newcommand{\beq}{\begin{equation}}
\newcommand{\eeq}{\end{equation}}
\newcommand{\bea}{\begin{eqnarray}}
\newcommand{\eea}{\end{eqnarray}}
\newcommand{\nn}{\nonumber}
\newcommand{\ra}{\rightarrow}
\newcommand{\DD}{{\cal D}}
\newcommand{\MSb}{$\overline{\mbox{MS}}$}
\newcommand{\as}{\alpha_{\rm s}}
\newcommand{\ars}{a_{\rm s}}
\newcommand{\ep}{\epsilon}

\def\z#1{{\zeta_{#1}}}
\def\zs{{\zeta_{2}^{\,2}}}

\def\ca{{C_{\!A}}}
\def\cas{{C^{\,2}_{\!A}}}

\def\cf{{C_F}}
\def\cfs{{C^{\, 2}_F}}
\def\cft{{C^{\, 3}_F}}
\def\nf{{n^{}_{\! f}}}
\def\nsq{{n^{\,2}_{\! f}}}

\begin{document}
\title{SUDAKOV RESUMMATIONS AT HIGHER ORDERS%
%
\thanks{Presented by S.M. and A.V. at the conferences `Matter to the 
Deepest', Ustron (Poland), September '05, and RADCOR 2005, Shonan Village
(Japan), October '05}%
}
\author{
S. Moch
\address{
Deutsches Elektronensynchrotron DESY\\ 
Platanenallee 6, D--15735 Zeuthen, Germany\\[4mm]}
A. Vogt
\address{
IPPP, Department of Physics, University of Durham\\
South Road, Durham DH1 3LE, United Kingdom\\[4mm]}
J. Vermaseren
\address{
NIKHEF,  
Kruislaan 409, 1098 SJ Amsterdam, The Netherlands\\[-6mm]}
}
\maketitle
\begin{abstract}
\noindent
We summarize our recent results on the resummation of hard-scattering
coefficient functions and on-shell form factors in massless perturbative QCD.
The threshold resummation has been extended to the fourth logarithmic order
for deep-inelastic scattering, Drell-Yan lepton pair production and
Higgs production via gluon-gluon fusion. The leading six infrared pole terms
have been derived to all orders in the strong coupling constant for 
the photon-quark-quark and the (heavy-top) Higgs-gluon-gluon form factors.
These results have many implications, most notably they lead to a new best
estimate for the Higgs production cross section at the LHC. 
\end{abstract}
%
%
\section{Introduction}
%
%
Coefficient functions, or partonic cross sections, form the backbone of
perturbative QCD. These quantities are calculable as a power series in the 
strong coupling constant $\as$, but exhibit large logarithmic corrections close
to threshold. The all-order resummation of the dominant soft-gluon 
contributions takes the form of an exponentiation in Mellin-$N$ space~\cite
{Sterman:1987aj,Catani:1989ne,Catani:1996yz,Contopanagos:1996nh}, 
where the moments $N$ are defined with respect to the appropriate scaling 
variable, like Bjorken-$x$ in deep-inelastic scattering (DIS) and $x = 
M_{\,l^+l^-\! ,\,H} ^{\,2}/s$ for the Drell-Yan (DY) process and Higgs 
production via gluon-gluon fusion.

\vspace{0.8mm}
The purpose of the exponentiation is (at least) two-fold. On the one hand, it 
can directly lead to improved phenomenological predictions close to exceptional 
kinematic points, for instance to an improved stability under scale variations.
On the other hand, it can be viewed as a generating functional of fixed-order 
perturbation theory close to the partonic thresholds. Hence progress in the 
soft-gluon resummation also facilitates improved fixed-order predictions which, 
depending on the specific observable, can be relevant even very far from the 
hadronic threshold. 

\vspace{0.8mm}
In this contribution we discuss recent results for the threshold resummation 
up to the fourth logarithmic (N$^3$LL) order~\cite{Moch:2005ba,Moch:2005ky}, 
and briefly illustrate their implications. We also summarize our recent results 
\cite{Moch:2005id,Moch:2005tm} for the on-shell quark and gluon form factors 
and their exponentiation~\cite{Collins:1980ih,Sen:1981sd,Magnea:1990zb,%
Ravindran:2004mb}, which were instrumental in extending the soft-gluon 
resummation to N$^3$LL accuracy for lepton-pair and Higgs boson production. 
Moreover the form-factor results are interesting also in a wider context, e.g.,
they provide another link to recent calculations performed in ${\cal N}\!=\!4$ 
Super-Yang-Mills theory~\cite{Bern:2005iz}.
%
%
\section{General structure of the threshold resummation}
%
%
As mentioned in the introduction, the coefficient functions for inclusive DIS,
Drell-Yan lepton-pair production and Higgs boson production exponentiate after
transformation to Mellin $N$-space~\cite{Sterman:1987aj,Catani:1989ne},
\beq
\label{eq:cNres}
  C^{N} \; =\; ( 1 + \ars\,g_{01}^{} + \ars^{\,2}\,g_{02}^{} + \ldots )
  \cdot \exp\, (G^N) \: + \: {\cal O}(N^{-1}\ln^n N) \:\: . 
\eeq
Here $g_{0k}^{}$ collects the $N$-independent contributions at $k$-th order in 
the strong coupling constant $\as$. The resummation exponent $G^N$ contains 
terms of the form $\ln^{k} N$ to all orders in $\as$ and takes the form
\beq
\label{eq:GNexp}
  G^N \: = \:
  \ln N \cdot g_1^{}(\lambda) \, + \, g_2^{}(\lambda) \, + \,
  \ars\, g_3^{}(\lambda) \, + \, \ars^{\,2}\, g_4^{}(\lambda) \, + \,
  \ldots 
\eeq
with $\lambda = \beta_0\, \ars\:\! \ln N$. The functions $g_k^{}$ represents 
the contributions of the $k$-th logarithmic (N$^{k-1}$LL) order. All our
relations refer to the \MSb\ scheme. 

\vspace{0.8mm}
The exponential in Eq.~(\ref{eq:cNres}) is build up from universal 
radiative factors $\Delta_{\,\rm p}$ and $J_{\rm p}$ due to radiation collinear
to the  initial- and final-state partons, and a process-dependent contribution 
$\Delta^{\rm int}$ from large-angle soft gluons. 
For example, the resummation exponents for the processes considered here read
\bea
\label{eq:GNdec}
  G_{\rm DIS}^N \quad\; & = & \,
    \ln \Delta_{\:\!\rm q} \: + \: \ln J_{\rm q} \: + \: 
    \ln \Delta^{\:\!\rm int}_{\:\!\rm DIS} \:\: , \nn \\
  G_{\rm \{DY,H\}}^N    & = & \,
    2\, \ln \Delta_{\:\!\rm \{q,g\}} \: + \: 
    \ln \Delta^{\:\!\rm int}_{\:\!\rm \{DY,H\}} \:\: .
\eea
$\Delta_{\:\!\rm p}$, the so-called jet function $J_{\rm p}$ and 
$\Delta^{\rm int}$ are given by certain integrals over functions of the running
coupling, $A_{\rm p}$, $B_{\rm p}$ and $D$. Specifically, the functional 
dependences are $\Delta_{\:\!\rm p}(A_{\rm p})$, $J_{\rm p}(A_{\rm p}, 
B_{\rm p})$ and $\Delta^{\rm int}(D)$. The functions $A_{\rm p}$, $B_{\rm p}$ 
and $D$, in turn, are defined in terms of power expansions in $\as$, for which 
we generally employ the convention 
\beq
\label{eq:fexp}
  f(\as) \:\: = \:\: 
  \sum_{k=1}^{\infty}\, f_k\, \left({\as \over 4\pi}\right)^k \:\: \equiv \:\: 
  \sum_{k=1}^{\infty}\, f_k\, \ars^{\,k} \:\: .
\eeq
The extent to which these functions are known sets the accuracy to which the 
threshold logarithms can be resummed.
It is worth noting that the function $D_{\:\!\rm DIS}^{}$ is found to vanish to 
all orders~\cite{Forte:2002ni,Gardi:2002xm}, hence 
$\Delta_{\:\!\rm DIS}^{\:\!\rm int} \; = \; 1$. 

\vspace{0.8mm}
The explicit expressions for the functions $g_i^{}(\lambda)$ in Eq.~(\ref
{eq:GNexp}) are obtained by performing the above-mentioned integrations, for 
instance using properties of harmonic sums and algorithms for the evaluation of 
nested sums \cite{Vermaseren:1998uu,Blumlein:1998if,Moch:2001zr,Moch:2005uc}. 
Specifically, $g_3^{}$ and $g_4^{}$ have been determined in Refs.~\cite
{Vogt:2000ci,Catani:2003zt} and \cite{Moch:2005ba}, to which the reader is 
referred for details. While the leading-log (LL) function $g_1^{}$
depends only on $A_1$, the N$^{k\geq1}$LL functions $g_{k+1}^{}$ include all 
parameters up to $A_{k+1}$, $B_k$ and $D_k$. We now turn to the present status 
of their determination.
%
%
\section{The known resummation coefficients}
%
%
The functions $A_{\rm p}$ are given by the leading large-$N$ (or large-$x$) 
coefficients of the diagonal splitting functions for the parton evolution, 
\beq
\label{eq:Pppdef}
  P_{\rm pp}(\as) \; = \; 
      A_{\rm p}(\as)\: (1-x)^{-1}_+
    + P_{\rm p}^{\,\delta}(\as)\: \delta(1-x)
    + {\cal O} \left( \ln (1-x) \right)
  \:\: ,
\eeq
which in turn are identical to the anomalous dimension of a Wilson line
with a cusp~\cite{Korchemsky:1989si}. The known expansion coefficients for the 
quark case read~\cite{Kodaira:1981nh,Moch:2004pa}
\bea
\label{eq:Aqexp}
  A_{\rm q,1} & \! = \! & \; 4\, \cf \:\: ,\nn \\[0.5mm]
  A_{\rm q,2} & \! = \! & \; 8\, \cf \left[ \left( \frac{67}{18} 
     - \zeta_2^{} \!\right) \ca - \frac{5}{9}\,\nf \right] \:\: ,\nn \\[1mm]
  A_{\rm q,3} & \! = \! &
     16\, \cf \left[ \cas \left( \frac{245}{24} - \frac{67}{9}
     \, \z2 + \frac{11}{6}\,\z3 + \frac{11}{5}\,\zs\! \right) 
   \, - \, \cf \nf \left( \frac{55}{24}  - 2\,\z3
   \!\right) \right. \nn\\ & & \left. \mbox{} \qquad
   + \: \ca \nf \left( - \frac{209}{108}
         + \frac{10}{9}\,\z2 - \frac{7}{3}\,\z3 \!\right)
   \, + \, \nsq \left( - \frac{1}{27}\right) \right] 
\eea
for $\nf$ effectively massless quark flavours. Here $\cf$ and $\ca$ are the 
usual colour factors ($\cf = 4/3$, $\ca = 3$ in QCD), and Riemann's zeta 
function is denoted by $\zeta_n$. The gluonic coefficients are related to 
Eqs.~(\ref{eq:Aqexp}) by~\cite{Korchemsky:1989si,Vogt:2004mw} 
\beq
\label{eq:Agexp}
A_{\rm g,i} \; = \; \ca / \cf \; A_{\rm q,i}\; .
\eeq
It is worthwhile to note that the $\zs$ terms in $A_{\rm p,3}$ have been 
confirmed by the recent ${\cal N}\!=\!4$ Super-Yang-Mills (SYM) calculation of 
Ref.~\cite{Bern:2005iz}.

\vspace{0.8mm}
The perturbative expansion of the functions $A_{\rm p}(\as)$ is very benign. 
In fact, already $A_3$ has a very small effect on the resummed coefficient
functions \cite{Vogt:2000ci,Catani:2003zt}. 
Therefore it is sufficient to estimate the presently unknown fourth-order 
coefficients $A_4$ entering $g_4^{}$ by their [1/1] Pad\'e approximants,
\beq
\label{eq:A4pade}
  A_{\rm q,4} \; \approx \; 7849\: ,\:\: 4313\: ,\:\:  1553 \quad
  \mbox{for} \quad \nf \; = \; 3\: ,\:\: 4\: ,\:\: 5 \:\: ,
\eeq
to which we assign a conservative 50\% uncertainty in numerical 
applications. Eqs.~(\ref{eq:Aqexp}) and (\ref{eq:A4pade}) lead to the numerical
four-flavour expansion 
\beq
\label{eq:Aqnum}
  A_{\rm q}(\as,\nf\!=\! 4) \; \cong \; 0.4244\:\as \:
  ( 1 \: + \: 0.6381\: \as \: + \: 0.5100\: \as^2 \:
  + \:0.4_{\,[1/1]}\: \as^{3}\: + \: \ldots )
  \:\: .
\eeq

We now turn to the coefficients $B_{\rm p}$ entering the jet functions 
$J_{\rm p}$. These quantities can be determined by comparing the $\as$-%
expansion of Eqs.~(\ref{eq:cNres}) and (\ref{eq:GNexp}) with the results
of fixed-order calculations of the DIS coefficient functions, which we have
recently extended to the third order in $\as$~\cite{Vermaseren:2005qc}:
\bea
\label{eq:Bqexp}
  B_{\rm q,1} & \! =\!& - 3\: \cf 
\:\: ,
\nn\\[1mm]
  B_{\rm q,2}  &\! =\!&
  \cfs \* \left[ - {3 \over 2} + 12\: \* \z2 - 24\: \* \z3 \right]
  + \cf \* \ca \* \left[ - {3155 \over 54} + {44 \over 3}\: \* \z2 
  + 40\: \* \z3 \right] 
  \nn \\ & & \mbox{\hspn}
  + \cf \* \nf \* \left[ {247 \over 27} - {8 \over 3}\: \* \z2 \right]
\:\: ,
\nn\\[1mm]
  B_{\rm q,3}  &\! =\! & 
       \cft  \*  \left[  - {29 \over 2}\: - 18\: \* \z2 - 68\: \* \z3 
         - {288 \over 5}\: \* \zs + 32\: \* \z2 \* \z3 
         + 240\: \* \z5 \right]
\nonumber\\[0.5mm]
&&\mbox{\hspn}
       + \ca \* \cfs  \*  \left[  - 46 + 287\: \* \z2 
         - {712 \over 3}\: \* \z3 - {272 \over 5}\: \* \zs 
         - 16\: \* \z2 \* \z3 - 120\: \* \z5 \right]
\nonumber\\[0.5mm]
&&\mbox{\hspn}
       - \cas \* \cf  \*  \left[    {599375 \over 729} 
         - {32126 \over 81}\:\! \* \z2 - {21032 \over 27}\:\! \* \z3 
         + {652 \over 15}\:\! \* \zs 
         \right.
         \left.
         + {176 \over 3}\:\! \* \z2 \* \z3 
         + 232\:\! \* \z5 \right]
\nonumber\\[0.5mm]
&&\mbox{\hspn}
       + \cfs \* \nf  \*  \left[ {5501 \over 54} - 50\: \* \z2 
         + {32 \over 9}\: \* \z3 \right]
       + \cf \* \nsq  \*  \left[  - {8714 \over 729} 
         + {232 \over 27}\: \* \z2 - {32 \over 27}\: \* \z3 \right]
\nonumber\\[0.5mm]
&&\mbox{\hspn}
       + \ca \* \cf \* \nf  \*  \left[ {160906 \over 729} 
         - {9920 \over 81}\: \* \z2 - {776 \over 9}\: \* \z3 
         + {208 \over 15}\: \* \zs \right]
\:\: .
\eea
The result for $B_{\rm q,1}$ is, of course, well-known~\cite{Sterman:1987aj,%
Catani:1989ne}, and $B_{\rm q,2}$ has been derived by us before in Ref.~\cite{
Moch:2002sn} where we explicitly established also  $D_2^{\,\rm DIS} = 0$. 
For the extraction of $B_{\rm q,3}$ \cite{Moch:2005ba}, on the other hand, we 
rely on the all-order proofs~\cite{Forte:2002ni,Gardi:2002xm} of $D_{\,\rm DIS}
= 0$ mentioned above.

\begin{figure}[bht]
\centerline{\epsfig{file=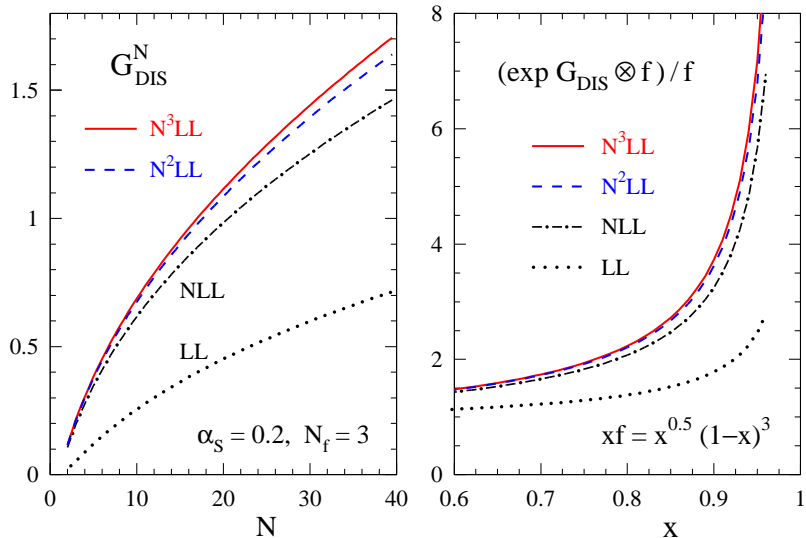,width=11.0cm,angle=0}}
\caption{\label{fig:GDIS}
Left: successive approximations for the resummation exponent (\ref{eq:GNexp})
of inclu\-sive DIS. Right: minimal-prescription~\cite{Catani:1996yz} 
convolutions with a typical input shape.}
\vspace*{-2mm}
\end{figure}

\vspace{1mm}
The numerical expansion of $B_{\rm q}$ in QCD is far less stable than 
Eq.~(\ref{eq:Aqnum}),
\beq
\label{eq:Bqnum}
  B_{\rm q}(\as,\nf\!=\! 4) \; \cong \; - 0.3183\,\as \,
  ( 1 \, - \, 1.227\, \as \, - \, 3.405\, \as^2 \, + \, \ldots )
  \:\: .
\eeq
Note, however, that the large third-order contribution to $B_{\rm q}$ actually
stabilizes the expansion of $G^N$ shown in Fig~\ref{fig:GDIS}: for $B_{\rm q,3}
=0$ and $N=40$, for example, the N$^3$LL term would be about as large as the 
previous order. 

\vspace{1mm}
The coefficients $B_{\rm g,i}$ for the gluonic jet function $J_{\,\rm g}$ are, 
for instance, relevant in direct-photon production which is dominated by the 
$\, {\rm q}\bar{{\rm q}}\ra {\rm g}\gamma\, $ and $\,{\rm qg}\ra {\rm q}
\gamma\,$ subprocesses close to threshold, see Ref.~\cite{Catani:1998tm}.
These coefficients can be obtained in the same manner as Eqs.~(\ref{eq:Bqexp}),
but from DIS by exchange of a scalar $\phi$ with a pointlike coupling to 
gluons, like the Higgs boson in limit of a heavy top quark. 
We have derived the corresponding coefficient function $C_{\phi,\rm DIS}$ 
up to the third order in the course of calculating the lower row of the 
flavour-singlet splitting function matrix~\cite{Vogt:2004mw}.
Comparison of these results to the expansion of Eq.~(\ref{eq:cNres}) yields 
$B_{\rm g,1}$ and the previously unknown quantities $B_{\rm g,2}$ and 
$B_{\rm g,3}$. The analytic results can be found in Ref.~\cite{Moch:2005ba}. 
Here we confine ourselves to the numerical expansion in four-flavour QCD,   
\beq
\label{eq:Bgnum}
  B_{\rm g}(\as,\nf\!=\! 4) \; \cong \; - 0.6631\,\as \,
  ( 1 \, - \, 0.7651\, \as \, - \, 2.696\, \as^2 \, + \, \ldots ) \:\: ,
\eeq
which shows a third-order enhancement similar to that in Eq.~(\ref{eq:Bqnum}).

\vspace{1mm}
Finally we address the process-dependent coefficients $D_{\:\!\rm i}$ due to 
the large-angle emission of soft gluons. Up to now, the two-loop coefficient 
functions for proton-proton processes are known only for the Drell-Yan cross 
section and Higgs boson production in the heavy-top approximation \cite
{Hamberg:1991np,Harlander:2001is,Anastasiou:2002yz,Ravindran:2003um}.
The corresponding coefficients $D_{\:\!2}^{\rm \{DY,H\}}$ have been extracted 
from these results in Refs.~\cite {Vogt:2000ci,Catani:2003zt}. 
Even for these processes, the three-loop coefficient functions have not been 
calculated so far. It is possible, however, to derive their third-order 
coefficients $D_3$ from mass-factorization constraints \cite{Moch:2005ky}, 
using our recent results for the pole terms of the three-loop quark and gluon 
form factors \cite{Moch:2005id,Moch:2005tm} and the third-order splitting 
functions \cite{Moch:2004pa,Vogt:2004mw}. Postponing the discussion of this 
derivation to section 5, the results for DY case read
\bea
\label{eq:DDYexp}
  D_1^{\,\rm DY} & = & 0 \:\: ,
\nn \\[0.5mm] 
  D_2^{\,\rm DY} & = & 
   \cf \left[ 
   \ca \left( - \frac{1616}{27} + \frac{176}{3}\,\z2 + 56\,\z3 \right)
   \: + \: \nf \left( \frac{224}{27} - \frac{32}{3}\,\z2 \right) 
   \right] \:\: ,
\nn \\[0.5mm] 
  D_3^{\,\rm DY} & = & 
      \cf \cas \left[ - {594058 \over 729} + {98224 \over 81}\,\z2
                   + {40144 \over 27}\,\z3 - {2992 \over 15}\,\zs
                   - {352 \over 3}\,\z2\z3 
                 \right.
\nn \\[0.5mm] & & \mbox{}
    - 384\,\z5 \biggr]         
    + \cf \ca\nf  \left[ {125252 \over 729} - {29392 \over 81}\,\z2
                   - {2480 \over 9}\,\z3 + {736 \over 15}\,\zs \right]
\nn\\[0.5mm] & & \mbox{\hspace*{-4mm}}
    + \cfs \nf \left[ {3422 \over 27} - 32\,\z2
                   - {608 \over 9}\,\z3 - {64 \over 5}\,\zs \right]
\nn\\[0.5mm] & & \mbox{\hspace*{-4mm}}
    + \cf \nsq \left[ - {3712 \over 729} + {640 \over 27} \,\z2
                   + {320 \over 27}\,\z3 \right] \:\: .
\eea
The corresponding coefficients for Higgs boson production via gluon-gluon
fusion are found to be related to these results by a simple colour-factor
substitution,
\beq
\label{eq:DHexp}
D_{\:\!\rm i}^{\,\rm H} \; = \; \ca / \cf \; D_{\:\!\rm i}^{\,\rm DY}\; ,
\eeq
which is in complete analogy to Eq.~(\ref{eq:Agexp}).
It worth pointing out that both the cusp anomalous dimensions $A_{\rm p}$ and 
the coefficients $D^{\:\!\rm DY}$ and $D^{\:\!\rm H}$ exhibit a maximally 
non-abelian colour structure, as anticipated for $A_{\rm p}$ in Ref.~\cite
{Korchemsky:1989si}.

\vspace{0.8mm}
The numerical expansion of $D^{\:\!\rm DY}$ in four-flavour QCD is given by
\beq
\label{eq:DDYnum}
  D^{\:\!\rm DY}(\as,\nf\!=\! 4) \; \cong \;  2.3211\,\as \,
  ( 0 \, + \, \as \: + \, 2.675\: \as^2 \, + \, \ldots ) \:\: .
\eeq
The ratio of the third- and second-order coefficients is very similar 
to that for the jet function in Eq.~(\ref{eq:Bqnum}), underlining the 
numerical relevance of $D_3$.
%
%
\section{On-shell form factors and their exponentiation}
%
%
The form factors of quarks and gluons are gauge invariant (but infrared 
divergent) parts of the perturbative corrections to inclusive hard scattering 
processes. They summarize the QCD corrections to the $qqX$ and $ggX$ vertices 
with a colour-neutral particle $X$ of either space-like or time-like momentum 
$q$. These quantities are also key ingredients in the infrared factorization
of general higher-order amplitudes~\cite{Catani:1998bh,Sterman:2002qn}.  

\vspace{0.8mm}
The relevant amplitude for the space-like $\gamma^{\ast}\!qq$ case is
\beq
\label{eq:fmudeq}
\Gamma_\mu  \: = \: {\rm i} e_{\rm q}\,
\bigl({\bar u}\, \gamma_{\mu\,} u\bigr)\, {\cal F}_{\rm q} (\as,Q^2)\; ,
\eeq
where $e_{\rm q}$ represents the quark charge and $Q^2 = -q^2$ the virtuality 
of the photon.  The gauge-invariant scalar function ${\cal F}_{\rm q}$ is the 
space-like quark form factor which can be calculated order by order in the 
strong coupling in dimensional regularization with $D=4-2\epsilon$.
The corresponding $H\:\!\!gg$ vertex defining ${\cal F}_{\rm g}$ is an effective 
interaction in the limit of a heavy top quark, 
\beq
\label{eq:LggH} 
   {\cal L}_{\rm eff} \; = \; - \frac{1}{4} \, C_H \, H \, 
   G^{\,a}_{\mu\nu} G^{\,a,\mu\nu} \:\: ,
\eeq
where $G^{\,a}_{\mu\nu}$ denotes the gluon field strength tensor, and the 
prefactor $C_H$ includes all QCD corrections, known to N$^3$LO~\cite
{Chetyrkin:1997un}, to the top-quark loop.

\vspace{0.8mm}
The well-known exponentiation of the form factors ${\cal F}$ is achieved by 
solving the evolution equations~\cite{Collins:1980ih,Sen:1981sd,Magnea:1990zb}
\beq
\label{eq:ffdeq}
Q^2 {\partial \over \partial Q^2} \ln {\cal F}\!\left(\as, {Q^2 \over \mu^2}, 
\epsilon\right) \:  = \:  
  {1 \over 2} \: K(\as,\epsilon) 
+ {1 \over 2} \: G\left({Q^2 \over \mu^2},\as,\epsilon \right) 
\eeq
based on a factorization of the form factor ${\cal F}$ into two functions $K$ 
and $G$.
The latter are subject to renormalization group equations~\cite{Collins:1980ih} 
which are both governed by the same anomalous dimension $A_{\rm p}$ of Eqs.~%
(\ref {eq:Aqexp}) and (\ref{eq:Agexp}) because, obviously, the sum of $G$ and 
$K$ in Eq.~(\ref{eq:ffdeq}) is a renormalization-group invariant. We follow the
decomposition of Refs.~\cite{Magnea:1990zb,Magnea:2000ss}, where the function 
$K$ is a pure counter-term collecting the infrared $1/\epsilon$ poles, while 
the infrared-finite function $G$ includes all dependence on the scale $Q^2$.

\vspace{0.8mm}
The resummed form factor is given as a double 
integral with the boundary condition ${\cal F}(\alpha_s,0,\epsilon) = 1$~\cite
{Magnea:1990zb}. After both integrations are performed, $\ln {\cal F}$ exhibits 
double logarithms of $Q^2/\mu^2$ and double poles in $\epsilon$. The relation 
(\ref{eq:ffdeq}) can be then used for a finite-order expansion and matching 
of the predictions to the results of explicit higher-order calculations. The
resulting expressions for the bare expansion coefficients ${\cal F}_{\rm i}$
in terms of the quantities $A_{\rm i}$ and the (still $\epsilon$-dependent) 
$\as$-expansion coefficients $G_{\rm i}$ of $G(Q^2/\mu^2=1)$ in Eq.~(\ref
{eq:ffdeq}) are sketched below (see Ref.~\cite{Moch:2005id} for the complete 
formulae): 
\bea
\label{eq:ffexp}
  {\cal F}_1 &\! =\! &
          - {1 \over 2 \ep^2} \* A_1
          - {1 \over 2\ep} \* G_1
\nn \\[0.3mm]
  {\cal F}_2 &\!\! =\!\! &
            {1 \over 8 \ep^4} \* A_1^2
          + {1 \over 8 \ep^3} \* A_1 \* (
            2 \* G_1
          - \beta_0
          )
          + {1 \over 8 \ep^2} \* (
            G_1^2
          + \ldots
          - A_2
          )
          - {1 \over 4 \ep} \* G_2
\nn \\[0.3mm]
  {\cal F}_3 &\!\! =\!\! &
          - {1 \over 48 \ep^6} \* A_1^3
          + \ldots 
          + {1 \over 72 \ep^2} \* (
            9 \* G_1 \* G_2
          + \ldots
          - 4 \* A_3
          )
          - {1 \over 6 \ep} \* G_3
\nn \\[0.3mm]
  {\cal F}_4 &\!\! =\!\! &
            {1 \over 384 \ep^8} \* A_1^4
          + \ldots
          + {1 \over 96 \ep^2} \* (
            3 \* G_2^2
          + 8 \* G_1 \* G_3
          + \ldots
          - 3 \* A_4
          )
          - {1 \over 8 \ep} \* G_4 \:\: . \;\;
\eea

We have extracted all three-loop pole terms of the quark and gluon form factors
${\cal F}_{\rm \!q}$ and ${\cal F}_{\rm \!g}$ from the calculation of the 
third-order coefficient functions for DIS by the exchange of a photon (coupling
to quarks) and a scalar $\phi$ (coupling to gluons)~\cite{Vermaseren:2005qc}, 
already mentioned above in the discussion of the jet function $J_{\rm p}$.
The details will be reviewed in the next section.

\noindent
Similar to the two-loop analysis of Ref.~\cite{Ravindran:2004mb}, we write
the coefficients $G_{\rm p}$ as
\bea
\label{eq:gf}
  G_{\rm p, 1} & = & 2 \left( P_{\rm p,\,1}^{\,\delta} 
    - \:\delta_{\rm pg} \beta_0 \right) \:\:
     + f_1^{\,\rm p} + \ep \widetilde{G_{\rm p, 1}} 
\:\: , \nn \\
  G_{\rm p, 2} & = & 2 \left( P_{\rm p,\,2}^{\,\delta} 
    - 2 \delta_{\rm pg} \beta_1 \right)
     + f_2^{\,\rm p} + \beta_0 \widetilde{G_{\rm p, 1}}(\ep\!=\!0) 
     + \ep \widetilde{G_{\rm p, 2}} 
\:\: , \nn \\
  G_{\rm p, 3} & = & 2 \left( P_{\rm p,\,3}^{\,\delta} 
    - 3 \delta_{\rm pg} \beta_2 \right)
     + f_3^{\,\rm p} + \beta_1 \widetilde{G_{\rm p, 1}}(\ep\!=\!0) 
  \nn \\ & & \mbox{\hspn}
     + \beta_0 \Big[ \widetilde{G_{\rm p, 2}}(\ep\!=\!0)
     - \beta_0\widetilde{\widetilde{G_{\rm p, 1}}}(\ep\!=\!0) \Big]  
     + \ep \widetilde{G_{\rm p, 3}} 
\eea
with 
$ \widetilde{F} = \ep^{-1} \left[ \, F - F (\ep\!=\!0) \, \right] $.
The quantities $P_{\rm p}^{\,\delta}$ have been defined in Eq.~(\ref
{eq:Pppdef}) above, and the terms with $\delta_{\rm pg}$ are due to the
renormalization of the operator $G_{\mu\nu} G^{\mu\nu}$ in Eq.~(\ref{eq:LggH}).
The crucial point of the decomposition~(\ref{eq:gf}) is that the functions
$f_i^{\,\rm p}$ turn out to be universal and, like the $A_{\rm p}$ in 
Eqs.~(\ref{eq:Aqexp}) and (\ref{eq:Agexp}) maximally non-Abelian with
(at least up to the third order) 
\beq
\label{eq:fgexp}
 f_i^{\,\rm g} \; = \; \ca / \cf \; f_i^{\,\rm q}\; .
\eeq
The explicit results for the quark case read
\bea
\label{eq:fqexp}
  f_1^{\,\rm q} & = & 0 
\:\: , \quad  f_2^{\,\rm q} \; = \; 2 \cf \left\{ 
    - \beta_0 \z2 - \frac{56}{27}\, \nf + \ca \left( 
    \frac{404}{27} - 14 \z3 \right) \right\}
\:\: , \nn\\[1.5mm]
  f_3^{\,\rm q} & = & 
  \cf \cas \left(
     {136781 \over 729} - {12650 \over 81}\,\z2 - {1316 \over 3}\, \z3
   + {352 \over 5}\,\zs + {176 \over 3}\,\z2\z3 + 192\,\z5
  \!\right)
\nn \\ & & \mbox{\hspn}
+ \ca\cf\nf \left(
   - {11842 \over 729} 
   + {2828 \over 81}\,\z2 + {728 \over 27}\, \z3 - {96 \over 5}\,\zs 
   \!\right)
+ \cfs\nf \left(
   - {1711 \over 27} 
\right.
\nn \\ & & \left. \mbox{}
   + 4\,\z2 + {304 \over 9}\, \z3 + {32 \over 5}\,\zs 
  \!\right)
+ \cf\nsq \left(
   - {2080 \over 729} - {40 \over 27} \,\z2 + {112 \over 27}\, \z3
  \!\right)
\:\: .
\eea
Note that $f_2^{\,\rm q}$ has been obtained already in Ref.~\cite
{Ravindran:2004mb}, and that the coefficients of the highest $\zeta$-function
weights, $\z2\z3$ and $\z5$ at three loops, agree with the results inferred 
from the recent ${\cal N}\!=\!4\,$ SYM calculation in Ref.~\cite{Bern:2005iz}.

\vspace*{0.8mm}
Going back to Eq.~(\ref{eq:ffexp}), it is worth noting that the leading term 
of $G_3$ in Eq.~(\ref{eq:gf}), together with corresponding coefficients of 
$G_1$ and $G_2$ to higher powers in $\ep$ (see Refs.~\cite{Moch:2005id,%
Moch:2005tm} for the explicit results) fix the six highest poles of the form
factors at four loops and, in fact, at all higher orders. Moreover, taking
into account that the numerical effect of $A_4$ in Eq.~(\ref{eq:Aqnum}) is 
small, our present results are sufficient for deriving the infrared finite 
absolute ratio $|{\cal F}_{\rm p} (q^2) / {\cal F}_{\rm p} (-q^2)|^2$ of the 
time-like and space-like form factors up to the fourth order in $\as$. 
The corresponding numerical results for $\nf = 4,\,5$ read
\bea
 & & \nn \\[-5mm]
 q\bar{q}\gamma^{\ast} &\! : & 
     1 + 2.094\, \as + 5.613\: \as^{\,2}
       + 15.70\, \as^{\,3} + (48.63 \pm 0.43)\, \as^{\,4} 
 \; , \nn \\
 ggH &\! : & 
     1 + 4.712\, \as + 13.69\, \as^{\,2}
       + 25.94\, \as^{\,3} + (36.65 \pm 0.35)\, \as^{\,4} 
 \; ,
\eea
where the the uncertainty of the last terms is due that of the fourth-order
cusp anomalous dimensions $A_{\rm p,4}$, as estimated below 
Eq.~(\ref{eq:A4pade}) in section 2.
%
%
\section{Partonic cross section and their infrared pole structure}
%
%
In this section, we finally discuss the extraction of the form factors from 
our calculation of the coefficient functions for inclusive DIS and the related 
derivation of all soft-enhanced third-order terms for the Drell-Yan process and
Higgs production, and thus of $D_3$ given already in Eqs.~(\ref{eq:DDYexp}) and
(\ref{eq:DHexp}), from these form-factor results and mass-factorization
constraints \cite{Moch:2005ky}.
  
\vspace*{0.8mm}
The starting points for the first step are the explicit results for the bare 
(unrenormalized and unfactorized) partonic structure functions $F^{\rm b}$ for 
$\,\gamma^{\ast} {\rm q} \ra {\rm q}X\,$ and $\,\phi^{\,\ast} {\rm g} \ra 
{\rm g}X\,$ in the limit $x \ra 1$ \cite{Vermaseren:2005qc}. At each order 
$\as^{\,n}$ keeping only the singular pieces proportional to $\delta(1-x)$ and 
the +-distributions 
\beq
\label{eq:DDdef}
  \DD_{\:\!l} \: = \: \left[ \frac{\ln^{\,l}(1-x)}{(1-x)} \right]_+ \:\: ,
  \quad\quad l \: = \: 1,\,\ldots\, 2n-1 \; ,
\eeq
these results are compared to the general structure of the 
$n$-th order contribution $F^{\rm b}_n$ in terms of the $l$-loop form factors 
${\cal F}_l$ and the corresponding real-emission parts ${\cal S}_{l\,}$,
\bea
F^{\rm b}_0
     &\,=\,& \delta(1-x) \nn \\[0.5mm]
F^{\rm b}_1
     &\,=\,& 2 {\cal F}_1\,\delta(1-x) + {\cal S}_1 \nn \\[0.5mm]
F^{\rm b}_2
     &\,=\,& \left(2 {\cal F}_2 + {\cal F}_1^{\,2} \right) \delta(1-x)
           + 2 {\cal F}_1 {\cal S}_1 + {\cal S}_2 \nn \\
F^{\rm b}_3
     &\,=\,& \left(2 {\cal F}_3 + 2 {\cal F}_1 {\cal F}_2\right) \delta(1-x)
           + \left(2 {\cal F}_2 + {\cal F}_1^{\,2} \right) {\cal S}_1 
           + 2 {\cal F}_1 {\cal S}_2 
           + {\cal S}_3\; . \quad
\label{eq:Fbdec}
\eea
In DIS the $x$-dependence of the real emission factors ${\cal S}_k$ is of the 
form ${\cal S}_k(f_{k,\ep})$, with the $D$-dimensional +-distributions 
$f_{k,\ep}$ defined by
\beq
\label{eq:Dplus}
  f_{k,\ep}(x) \; = \; \ep [\,(1-x)^{-1-k\ep}\,]_+
               \; = \; - {1 \over k}\, \delta(1-x) + \sum_{i=0}\,
                       {(-k \ep)^i \over i\, !}\,\ep\,\DD_{\:\!i}\; . 
\eeq
The dimensionally regularized (with $D=4-2\epsilon$) bare structure
functions $F^{\rm b}_n$ in Eq.~(\ref{eq:Fbdec}) exhibit poles in $\ep$ up to 
$\ep^{-2n}$, with a structure completely determined by mass factorization.
On the other hand, the individual real and virtual contributions ${\cal F}_k$ 
and ${\cal S}_k$ in Eq.~(\ref{eq:Fbdec}) contain poles up to order $\ep^{-2k}$, 
which cancel due to the Kinoshita--Lee-Nauenberg theorem~\cite
{Kinoshita:1962xx,Lee:1964xx}.

\vspace*{0.8mm}
The determination of the form factor now proceeds as follows.
Once the combinations of lower-order quantities in Eq.~(\ref{eq:Fbdec}) have 
been subtracted from $F^{\rm b}_n$, the $n$-loop form factor ${\cal F}_n$ can 
simply be extracted by the substitution
\beq
\label{eq:DD0subs}
  \DD_{\:\!0} \: \to \: {1 \over n \ep} \, \delta(1 - x)
  - \sum_{i=1}\, {(-n \ep)^i  \over i\, !}\, \DD_{\:\!i} \; ,
\eeq
which exploits the particular analytical dependence of ${\cal S}_k$ on $x$,
i.e., Eq.~(\ref{eq:Dplus}). 
As $\delta(1-x)$ enters with a factor $1/\ep$, this extraction loses one power
in $\ep$. Hence from the third-order calculation to order $\ep^0$, as performed 
for the coefficient function, we can only extract all pole terms of 
${\cal F}_3$ in this manner.
   
\begin{figure}[thb]
\centerline{\epsfig{file=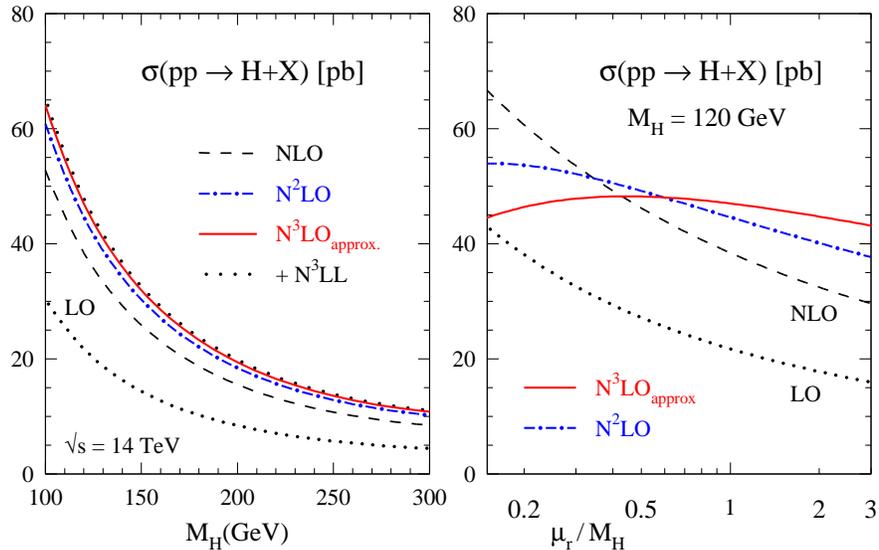,width=12.0cm,angle=0}}
\caption{\label{fig:HLHC}
The perturbative expansion of the total cross section for Higgs production
at the LHC. Left: dependence on the Higgs mass $M_{\rm H}$. 
Right: renormalization-scale (in-)$\:\!$stability for $M_{\rm H} = 
120\mbox{ GeV}$. See Ref.~\cite{Moch:2005ky} for a detailed discussion.} 
\vspace*{-1mm}
\end{figure}

\vspace*{0.8mm}
The second step, the determination of the +-distribution contributions to 
coefficient functions for lepton-pair and and Higgs boson production, proceeds 
along similar lines, see Ref.~\cite{Matsuura:1987wt} for an early two-loop 
application to the Drell-Yan process. In analogy to Eq.~(\ref{eq:Fbdec}), the 
soft limit of the bare partonic cross sections $W^{\rm b}$ for 
$\, {\rm q}\bar{{\rm q}} \,\ra\, \gamma^{\ast} \,\ra\, l^+ l^-\,$ and 
$\,{\rm gg} \,\ra\, H\,$ reads
\bea
\label{eq:Wbexp}
 W^{\rm b}_0
   &\,=\,& \delta(1-x) \nn \\[0.3mm]
 W^{\rm b}_1
   &\,=\,& 2\,\mbox{Re}\,{\cal F}_1\,\delta(1-x) + {\cal S}_1 \nn \\[0.3mm]
 W^{\rm b}_2
   &\,=\,& (2\,\mbox{Re}\,{\cal F}_2 + \left|{\cal F}_1\right|^2)\, \delta(1-x)
         + 2\,\mbox{Re}\, {\cal F}_1 {\cal S}_1 + {\cal S}_2 \nn\\[0.3mm]
 W^{\rm b}_3
   &\,=\,& (2\,\mbox{Re}\,{\cal F}_3 + 2 \left|{\cal F}_1 {\cal F}_2 \right|)
         \, \delta(1-x)
         + (2\,\mbox{Re}\,{\cal F}_2 + \left|{\cal F}_1\right|^2) {\cal S}_1 
\nn\\[0.3mm]
& & \mbox{}
         + 2\,\mbox{Re}\,{\cal F}_1 {\cal S}_2 + {\cal S}_3\; ,
\eea
where, of course, ${\cal F}$ now denotes the time-like quark or gluon form 
factor, known by analytic continuation from $q^2 = -Q^2 < 0\,$ to $q^2 > 0$.
The real-emission contributions ${\cal S}_k$ depend on the scaling variable 
$x = M_{\gamma^{\,\ast}\! ,\,H}^{\,2}/s$. In this case, the dependence of 
${\cal S}_k$ on $x$ is of the form ${\cal S}_k(f_{2k,\ep})$, i.e., 
\beq
\label{eq:SsoftDYH}
  {\cal S}_k \; = \; f_{2k,\ep}\, \sum_{l=-2k}^{\infty} 
  2k\, s_{k,l}^{}\: \ep^{\,l} \; .
\eeq
With the known time-like form factors, the expansion coefficients $s_{k,l}^{}$ 
of the soft function ${\cal S}_k$ can be derived recursively as far as they
are subject to the KLN cancellations and the mass-factorization structure
relating the remaining poles to the splitting functions (\ref{eq:Pppdef}). 
Employing the results of Refs.~\cite{Moch:2005id,Moch:2005tm} and \cite
{Moch:2004pa,Vogt:2004mw}, the third-order terms $s_{3,-6}^{}\,\ldots\,s_{3,-1}$
can be obtained. Due to Eq.~(\ref{eq:Dplus}) this is sufficient to derive all
+-distribution contributions to the third-order coefficient functions, in
particular also the coefficient of $\DD_0$ from which $D_3^{\{\rm DY,\rm H\}}$ 
in Eqs.~(\ref{eq:DDYexp}) and (\ref{eq:DHexp}), can be determined by matching.
An important application on these new results is presented in 
Fig.~\ref{fig:HLHC}.

\vspace*{0.8mm}
The connection between mass-factorization and resummation leads to a
simple relation between the coefficients $D_{\rm n}$ and 
$f_{\rm n}^{\:\!\rm p}$ in Eqs.~(\ref{eq:fgexp}) and (\ref{eq:fqexp}),
\bea
\label{eq:Dvsf}
  D_2^{\,\{ \rm DY,\,\rm H \}} & = & - 2 f_2 + 2\beta_0 s_{1,0}^{} 
\nn \\
  D_3^{\,\{ \rm DY,\,\rm H \}} & = & - 2 f_3 + 2\beta_1 s_{1,0}^{} 
  - 4\beta_0^2 s_{1,1}^{}
  + 4 \beta_0 \!\left[ s_{2,0}^{} 
    - {36 \over 5}\,\zs\,C_{\{F,A\}}^{\,2} \right]
  \; , \:\:
\eea
which has also been derived by extending the threshold resummation to the 
$N$-independent contributions~\cite{Eynck:2003fn,Laenen:2005uz}, see also
Ref.~\cite{Idilbi:2005ni}. In our approach, the $s_{n,l}^{}$ terms 
can be traced back to the $\as$-renormalization of Eqs.~(\ref{eq:Wbexp}).

%
%
\section{Summary}
%
%
Building on our third-order computation of the splitting functions
\cite{Moch:2004pa,Vogt:2004mw} and the coefficient functions for inclusive
DIS \cite{Vermaseren:2005qc}, we have derived new three-loop and all-order
results for the threshold resummation~\cite{Moch:2005ba,Moch:2005ky}, the 
on-shell quark and gluon form factors~\cite{Moch:2005id,Moch:2005tm}, and the
coefficient functions for lepton-pair and Higgs boson production at proton
colliders~\cite{Moch:2005ky}. These results have important implications within
and beyond perturbative QCD. 
\vspace*{-1mm}
\section*{Acknowledgments}
The work of S.M. has been supported in part by the Helmholtz Gemeinschaft 
under contract VH-NG-105 and by the Deutsche Forschungsgemeinschaft in 
Sonderforschungs\-be\-reich/Transregio 9.
The work of J.V. has been part of the research program of the Dutch 
Foundation for Fundamental Research of Matter (FOM).
%
%


\vspace{-2mm}
\begin{thebibliography}{10}
\vspace{-1mm}

\bibitem{Sterman:1987aj}
G. Sterman,
\newblock Nucl. Phys. B281 (1987) 310
\newblock 

\bibitem{Catani:1989ne}
S. Catani and L. Trentadue,
\newblock Nucl. Phys. B327 (1989) 323, B353 (1991) 183
\newblock 
\newblock 

\bibitem{Catani:1996yz}
S.$\,$Catani, M.$\,$Mangano, P.$\,$Nason, L.$\,$Trentadue,
\newblock Nucl.$\,$Phys.$\,$B478 (1996) 273
\newblock 

\bibitem{Contopanagos:1996nh}
H. Contopanagos, E. Laenen and G. Sterman,
\newblock Nucl. Phys. B484 (1997) 303
\newblock 

\bibitem{Moch:2005ba}
S. Moch, J.A.M. Vermaseren and A. Vogt,
\newblock Nucl. Phys. B726 (2005) 317
\newblock 

\bibitem{Moch:2005ky}
S. Moch and A. Vogt,
\newblock hep-ph/0508265 (Phys. Lett. B, in press)
\newblock 

\bibitem{Moch:2005id}
S. Moch, J.A.M. Vermaseren and A. Vogt,
\newblock JHEP 08 (2005) 049
\newblock 

\bibitem{Moch:2005tm}
S. Moch, J.A.M. Vermaseren and A. Vogt,
\newblock Phys. Lett. B625 (2005) 245
\newblock 

\bibitem{Collins:1980ih}
J.C. Collins,
\newblock Phys. Rev. D22 (1980) 1478 
\newblock 

\bibitem{Sen:1981sd}
A. Sen,
\newblock Phys. Rev. D24 (1981) 3281 
\newblock 

\bibitem{Magnea:1990zb}
L. Magnea and G. Sterman,
\newblock Phys. Rev. D42 (1990) 4222 
\newblock 

\bibitem{Ravindran:2004mb}
V. Ravindran, J. Smith and W.L. van Neerven,
\newblock Nucl. Phys. B704 (2005) 332
\newblock 

\bibitem{Bern:2005iz}
Z. Bern, L.J. Dixon and V.A. Smirnov,
\newblock  Phys. Rev. D72 (2005) 085001
\newblock 

\bibitem{Forte:2002ni}
S. Forte and G. Ridolfi,
\newblock Nucl. Phys. B650 (2003) 229
\newblock 

\bibitem{Gardi:2002xm}
E. Gardi and R.G. Roberts,
\newblock Nucl. Phys. B653 (2003) 227
\newblock 

\bibitem{Vermaseren:1998uu}
J.A.M. Vermaseren,
\newblock Int. J. Mod. Phys. A14 (1999) 2037
\newblock 

\bibitem{Blumlein:1998if}
J. Bl\"umlein and S. Kurth,
\newblock Phys. Rev. D60 (1999) 014018
\newblock 

\bibitem{Moch:2001zr}
S. Moch, P. Uwer and S. Weinzierl,
\newblock J. Math. Phys. 43 (2002) 3363
\newblock 

\bibitem{Moch:2005uc}
S. Moch and P. Uwer,
\newblock math-ph/0508008
\newblock 

\bibitem{Vogt:2000ci}
A. Vogt,
\newblock Phys. Lett. B497 (2001) 228
\newblock 

\bibitem{Catani:2003zt}
S. Catani, D.~de Florian, M.~Grazzini and P.~Nason,
\newblock JHEP 07 (2003) 028
\newblock 

\bibitem{Korchemsky:1989si}
G.P. Korchemsky,
\newblock Mod. Phys. Lett. A4 (1989) 1257
\newblock 

\bibitem{Kodaira:1981nh}
J. Kodaira and L. Trentadue,
\newblock Phys. Lett. B112 (1982) 66 
\newblock 

\bibitem{Moch:2004pa}
S. Moch, J.A.M. Vermaseren and A. Vogt,
\newblock Nucl. Phys. B688 (2004) 101
\newblock 

\bibitem{Vogt:2004mw}
A. Vogt, S. Moch and J.A.M. Vermaseren,
\newblock Nucl. Phys. B691 (2004) 129
\newblock 

\bibitem{Vermaseren:2005qc}
J.A.M. Vermaseren, A. Vogt and S. Moch,
\newblock Nucl. Phys. B724 (2005) 3
\newblock 

\bibitem{Moch:2002sn}
S. Moch, J.A.M. Vermaseren and A. Vogt,
\newblock Nucl. Phys. B646 (2002) 181
\newblock 

\bibitem{Catani:1998tm}
S. Catani, M.L. Mangano and P. Nason,
\newblock JHEP 9807 (1998) 024
\newblock  

\bibitem{Hamberg:1991np}
R. Hamberg, W. van Neerven and T. Matsuura,
\newblock Nucl. Phys. B359 (1991) 343, B644 (2002) 403 (E)
\newblock 

\bibitem{Harlander:2001is}
R.V. Harlander and W.B. Kilgore,
\newblock Phys. Rev. D64 (2001) 013015
\newblock 

\bibitem{Anastasiou:2002yz}
C. Anastasiou and K. Melnikov,
\newblock Nucl. Phys. B646 (2002) 220
\newblock 

\bibitem{Ravindran:2003um}
V. Ravindran, J. Smith and W.L. van Neerven,
\newblock Nucl. Phys. B665 (2003) 325
\newblock 

\bibitem{Catani:1998bh}
S. Catani,
\newblock Phys. Lett. B427 (1998) 161
\newblock 

\bibitem{Sterman:2002qn}
G.~Sterman and M.E. Tejeda-Yeomans,
\newblock Phys. Lett. B552 (2003) 48
\newblock 

\bibitem{Chetyrkin:1997un}
K.G.$\,$Chetyrkin, B.A.$\,$Kniehl and M.$\,$Steinhauser,
\newblock Nucl. Phys. B510 (1998) 61
\newblock 

\bibitem{Magnea:2000ss}
L. Magnea,
\newblock Nucl. Phys. B593 (2001) 269
\newblock 

\bibitem{Kinoshita:1962xx}
T. Kinoshita,
\newblock J. Math Phys. 3 (1962) 650
\newblock 

\bibitem{Lee:1964xx}
T.D. Lee and M. Nauenberg,
\newblock Phys. Rev. B133 (1964) 1549
\newblock 

\bibitem{Matsuura:1987wt}
T. Matsuura and W.L. van Neerven,
\newblock Z. Phys. C38 (1988) 623
\newblock 

\bibitem{Eynck:2003fn}
T.O. Eynck, E. Laenen and L. Magnea,
\newblock JHEP 06 (2003) 057
\newblock 

\bibitem{Laenen:2005uz}
E. Laenen and L. Magnea,
\newblock hep-ph/0508284 (Phys. Lett. B, in press)
\newblock 

\bibitem{Idilbi:2005ni}
A.~Idilbi, X.d.~Ji, J.P.~Ma and F.~Yuan,
\newblock hep-ph/0509294
\newblock 

\end{thebibliography}
\end{document}